
%
\magnification=\magstep1
\hsize=6.75truein
\vsize=8.7 truein
\baselineskip=16pt
\parskip=\medskipamount
\parindent= .35in
\def \di{\partial}
\def \lg {\tilde {\bf g}}
\def \lgp {\tilde {\bf g}_+}
\def \lgm {\tilde {\bf g}_-}
\def \LG {\tilde {\cal G}}

\def \ga {{\bf g}_{a,\kappa}}

\def \k {\kappa}
\def \eps {\epsilon}
\def \l {\lambda}
\def \o {\omega}

\def \tr{{\rm tr}}

\def \det{{\rm det}}
\def \Ad{{\rm Ad}}
\def \log {{\rm log}}

\def \ra {\rightarrow}
\def \lra{\longrightarrow}

\def \lmt {\longmapsto}
\def \wt {\widetilde}
\def \wh {\widehat}
\def \ss{\subset}

\def \Iak {{\cal I}_{a,\k}}
\def \scst {\scriptstyle}

\def\up#1{\leavevmode \raise .3ex\hbox{$#1$}}
\def\down#1{\leavevmode \lower .5ex\hbox{$\scriptstyle#1$}}
\def\chiJ{\up{\chi}_{\down{J_k}}}
\font\eightrm=cmr8
\font\eightbf=cmbx8
\font\eightit=cmti8
\font\grand=cmbx10 at 12pt
\def \smalltype{\let\rm=\eightrm  \let\bf=\eightbf
\let\it=\eightit \let\sl=\eightsl \let\mus=\eightmus
\baselineskip = 9.5pt minus .75pt  \rm}
\rightline{ CRM-1846 (1993)  \break}
\bigskip
\centerline {\grand Hamiltonian Structure of Equations}
\centerline{{\grand Appearing in Random Matrices}
\footnote{$^\dagger$}{\smalltype \baselineskip = 9.5pt minus 2pt
Research supported in part by the Natural Sciences and Engineering Research
Council of
Canada and by the National Science Foundation, U.S.A., DMS--9001794 and
DMS--9216203.}}
\bigskip
\centerline{J.~Harnad\footnote{$^1$}{\smalltype  e-mail address:
harnad@alcor.concordia.ca  {\it \ or\ }  harnad@mathcn.umontreal.ca}}
\centerline
{\smalltype \it Department of Mathematics and Statistics, Concordia
University}
\centerline{\smalltype \it  7141 Sherbrooke W., Montr\'eal, Canada H4B 1R6, \
and}
\centerline{\smalltype \it  Centre de recherches math\'ematiques,
Universit\'e de Montr\'eal}
\centerline{\smalltype \it  C.~P. 6128-A, Montr\'eal, Canada
H3C 3J7}
\medskip
\centerline {C.~A.~Tracy\footnote{$^2$}{\smalltype e-mail address:
catracy@ucdavis.edu}}
\centerline {\smalltype \it Department of Mathematics and
Institute of  Theoretical Dynamics}
\centerline{\smalltype \it University of California, Davis, CA 95616, USA}
\medskip
\centerline{H.~Widom\footnote{$^3$}{\smalltype  e-mail
address:widom@cats.ucsc.edu}}
\centerline{\smalltype \it  Department of
Mathematics}
\centerline{\smalltype \it  University of California, Santa Cruz, CA 95064,
USA.}
 \smallskip
\baselineskip 14pt
  \bigskip
{\smalltype \centerline{\bf Abstract}
\smallskip
 The level spacing distributions in the Gaussian Unitary
Ensemble,
 both
in the ``bulk of the spectrum,'' given by the Fredholm determinant of the
operator
with the {\it sine kernel\/} ${\sin \pi(x-y) \over \pi(x-y)}$ and on the
``edge of the spectrum,'' given by the {\it Airy kernel\/} ${\rm{Ai}(x)
\rm{Ai}'(y) -
\rm{Ai}(y) \rm{Ai}'(x) \over (x-y)}$, are determined by compatible systems
of nonautonomous Hamiltonian equations. These may be viewed as special
cases of isomonodromic deformation equations for first order ${\scst 2\times
2}$
matrix differential operators with regular singularities at finite points and
irregular
ones of Riemann index $ {\scst 1}$ or ${\scst 2}$ at ${\smalltype \infty}$.
Their
Hamiltonian structure is explained within the {\it classical ${\scst
R}$--matrix}
framework as the equations induced by spectral invariants on the loop algebra
${\scst \wt{sl}(2)}$, restricted to a Poisson subspace of its dual space
${\scst \wt{sl}^*_R(2)}$, consisting of elements that are rational in the loop
parameter.}  \vfill \eject

\noindent{\bf 0. Introduction}
\medskip
It has long been known   for a class of exactly solvable
models in statistical mechanics that  the correlation
functions are expressible in terms of  Painlev{\'e} transcendents or
more generally  solutions
to completely  integrable differential
equations possessing the Painlev{\'e} property {\bf [WMTB]},
{\bf [MTW]}, {\bf [SMJ]}. (For a review see {\bf [Mc]}.)\ \
It was the great insight of the Kyoto School to invent the notion of
a $\tau$--function associated to a system of  completely integrable
differential
equations {\bf [SMJ]}, {\bf [JMMS]}, {\bf [JMU]}, {\bf [JM]} and to realize
that in this class of solvable models the
  correlation functions are $\tau$--functions.
\par
In the theory of random matrices it has  also long been  known
(see, e.g., {\bf [Po]}, {\bf [Me1]}) that the
level spacing distributions are expressible in terms of the
Fredholm determinant of an integral operator.
  In a particular scaling limit the kernel of this
  integral operator is the
{\it sine kernel\/}.  For this  case
it was shown in
{\bf [JMMS]}
that the Fredholm determinant was also a $\tau$--function. This then
leads in the simplest  case to formulas for the level spacing distributions
in terms of a particular Painlev{\'e} V transcendent.  More generally,
the deformation equations consist of a compatible set of nonautonomous
Hamiltonian equations.
\par
   Since the original work of  {\bf [JMMS]} on random
matrices and impenetrable bosons,
  the methods for deducing the relevant
deformation equations and properties of the $\tau$--functions have been
simplified and generalized, both in the study of random matrices
{\bf [BTW]}, {\bf [Me2]}, {\bf [Dy]}, {\bf [MM]},
 {\bf [TW1]}, {\bf [TW2]}, {\bf [W]} and quantum correlation  functions {\bf
[IIKS]},
{\bf [IIKV]}. The Hamiltonian structure of the underlying
 deformation equations
can  most simply be understood as a nonautonomous version of the equations
induced by
spectral invariants on a loop algebra with respect to a suitably chosen {\it
classical
$R$--matrix} Poisson bracket structure {\bf [H]}.

	In the following section, the problem
 of computing level spacing distributions
for the Gaussian Unitary Ensemble (GUE), both
 in the ``bulk of the spectrum'' and at the
``edge of the spectrum'' is formulated in these terms---the two cases differing
only in the  kernels resulting from the
different  scaling limits. For the first case, this
is the
above mentioned  {\it sine kernel\/}
$${1\over \pi} {\sin \pi(x-y) \over x-y}\> , $$
 while for the second
it is the {\it Airy kernel\/}
$${\rm{Ai}(x) \rm{Ai}'(y) - \rm{Ai}(y) \rm{Ai}'(x)
\over x-y}\> .$$
In \S 2, the resulting Hamiltonian equations are explained in terms
of spectral invariants on the dual space $\wt{sl}^*(2)$ of the loop algebra
$\wt{sl}(2)$, with respect to a standard classical $R$--matrix Poisson bracket
structure, restricted to Poisson subspaces consisting of rational coadjoint
orbits.
\bigskip
\noindent{\bf 1. Spectral Distributions for Random Matrices}  \smallskip
\nobreak
 \noindent{\it 1a.\quad Fredholm Determinant Representation \hfill}

\nobreak
  The GUE measure on the spectral space of $N\times N$ hermitian  matrices
with eigenvalues $(X_1, \dots ,X_N)$
is {\bf [Me1]}
$$
P_{N }(X_1,\ldots,X_N) \,
dX_1\cdots dX_{N} = C_{N }\prod_{j<k}\left\vert
X_j-X_k\right\vert^2 \,  \exp\left(-\sum_j X_j^2\right) dX_1\cdots dX_N.
\eqno(1.1)
 $$
Averaging over all but $n$ of the eigenvalues gives the $n$--point correlation
function {\bf [Me1]}:
$$
\eqalign{
R_{n,N}(X_1,\ldots,X_n)
&={N!\over (N-n)! }
\int_{-\infty}^{\infty}\cdots\int_{-\infty}^{\infty}
 P_{N} (X_1,\ldots,X_N)\,
dX_{n+1}\cdots dX_N \cr
&=\det\left(K_N(X_j,X_k)
\Bigr\vert_{j,k=1}^n\right),} \eqno(1.2)
$$
where
$$
K_N(X,Y) :=\sum_{k=0}^{N-1} \varphi_k(X) \varphi_k(Y) \, . \eqno(1.3)
$$
Here $\lbrace \varphi_k(X) \rbrace$ is the sequence obtained
by orthonormalizing the sequence
$ \left\lbrace X^k \exp (-X^2/2) \right\rbrace $
over $(-\infty,\infty)$.
In particular $R_{1,N}(X)$, the density of eigenvalues at $X$, equals
$K_N(X,X)$.
\par
For fixed $X$,
$$R_{1,N}(X) \sim {1\over \pi} \sqrt{2N}\
\  {\rm as\ } \ N\rightarrow\infty . \eqno(1.4)$$
The  scaling limit in the bulk of the
spectrum  at the point $X_0$ is the limit
$$ N\rightarrow\infty,\  X_j\rightarrow X_0,\  {\rm such\ that\ \  }
x_j=(X_j-X_0)R_{1,N}(X_0) \ \ {\rm is\ fixed.}\eqno(1.5)$$
In this limit  the scaled n--point correlation functions
$R_n(x_1,\ldots , x_n)$, defined  by
$$R_{n}(x_1,\ldots , x_n) dx_1\cdots dx_n=
 \lim_{N\to\infty ,  X_j\to X_0 \atop
x_j {\ \rm fixed}} R_{n,N}(X_1,\ldots,
X_n)  dX_1\cdots dX_n  , \eqno(1.6)
$$
are given by {\bf [Me1]}
$$
R_{n}(x_1,\ldots,x_n)=\det\left(K(x_j,x_k)\Bigr\vert_{j,k=1}^n\right)
\eqno(1.7)
$$
where
$$
K(x,y) := {1\over \pi}\,  {\sin\pi(x-y)\over x-y}\> . \eqno(1.8)
$$
\par
   Now consider a set of ``spectral windows''
$$
 J=J_1\cup\cdots\cup  J_m
$$
where
$$
\{ J_k = [a_{2k-1},a_{2k}] \ss {\bf R}\}_{k=1, \dots m},
 \qquad a_1 < a_2 < \dots <
a_{2m}  \eqno(1.9)
$$
is an ordered set of disjoint intervals on the real axis.
The probability that a randomly chosen GUE matrix has
 precisely $\{n_1, n_2,
\dots, n_m\}$ scaled eigenvalues in the respective
 intervals $\{J_1, J_2, \dots ,
J_m\}$ is given by the {\it level spacing distribution}:
$$
E(n_1,\ldots,n_m,J) = {(-1)^{n}
\over n_1!\cdots n_m!}{\partial^{n} D(J;\lambda)\quad\over
\partial\lambda_1^{n_1}\cdots\partial\lambda_{m}^{n_m}}
\Bigr\vert_{\lambda_1=\cdots=\lambda_m=1} \eqno(1.10)
$$
with $n=n_1+\cdots+n_m$.
Here
$D(J;\lambda)$ is the Fredholm determinant
$$
D(J;\lambda):=\det(I - \wh K) \eqno(1.11)
$$
of the integral operator $\wh K$ with kernel
$$
K_J(x,y) :=\sum_{k=1}^m \lambda_k\,  K(x,y)\, \chiJ (y), \eqno(1.12)
$$
where $\chiJ (y)$ is the characteristic function for the interval $J_k$, and
$$
\l := (\l_1, \l_2, \dots , \l_m) \eqno(1.13)
$$
denotes the set of generating function parameters.
The case of one interval is in {\bf [Me1]}.  The more general formula (1.10)
was derived in {\bf [TW1]}.
\par
   In the following, we shall see that $D(J; \lambda)$
 is determined by a system of
partial differential equations consisting of a compatible set of nonautonomous
Hamiltonian equations defined on a suitable phase space. The quantity of
central
interest will be the total differential
$$
\o(a) :=d_{a}\log D(J; \l)=
 \sum_{i=1}^{2m}{\di \log D(J; \l)\over \di
a_i} \, da_i \eqno(1.14)
$$
of the logarithm of the generating function $D(J;\l)$, taken with respect to
the set of
parameters
$$
a := (a_1, \dots , a_{2m}) \eqno(1.15)
$$
giving the boundaries of the spectral domains. The dynamical equations of
Jimbo, Miwa,
M\^ori and Sato {\bf [JMMS]}, which determine the dependence of this quantity
on the
parameters $(a_1, \dots , a_{2m})$, are given below, following the approach
developed
in {\bf [TW1]}.
 We note that in
{\bf [JMMS]}, {\bf [TW1]}, and {\bf [TW2]} the case of equal
$\lambda_j$'s only was considered in the derivation of the equations.
   \smallskip
\noindent{\it 1b. Dynamics of Distribution Functions}

\nobreak
   The key object in what follows is the resolvent operator
$$
\wh R := (1 - \wh K)^{-1} \wh K . \eqno(1.16)
$$
Introducing the functions
$$
\eqalignno{
Q(x,a) &:= (1 - \wh K)^{-1}A  &(1.17a)\cr
P(x,a) &:= (1 - \wh K)^{-1}A^{\prime}, &(1.17b)}
$$
where
$$
 A(x) := {1\over \pi}{\rm sin}\pi x, \eqno(1.18)
$$
the kernel of the resolvent operator can be shown to be given by
$$
\eqalignno{
R(x,y) &= \sum_{k=1}^m \lambda_k{Q(x,a)P(y,a)-P(x,a) Q(y,a)\over x-y}
\chiJ (y)\, , \quad x\neq y, &(1.19a) \cr
 R(x,x) &=\sum_{k=1}^m \lambda_k\left({dQ\over dx}(x,a)\,  P(x,a)- {dP\over
dx}(x,a)\,
Q(x,a) \right)\chiJ (x). &(1.19b)}
 $$
Using the identity
$$
d_a\log\left(\det (I-\wh{K})\right) =
 - \tr\left((I-\wh{K})^{-1}d_a \wh{ K}\right), \eqno(1.20)
$$
it is easy to show that
$$
\o(a)= \sum_{j=1}^{2m}(-1)^{j+1} R(a_j,a_j)\, da_j,  \eqno(1.21)
$$
where $R(a_j,a_k)$ is obtained by
taking the limits $(x,y) \ra (a_j, a_k)$ inside $J$.
Let
$$
\eqalignno{
q_j&:=\lim_{x\rightarrow a_j \atop x\in J}\sqrt{\l_j}Q(x,a)  &(1.22a)\cr
 p_j&: = \lim_{x\rightarrow a_j \atop x\in J}\sqrt{\l_j}P(x,a),\ \
j=1,\ldots,2m.&(1.22b)}
$$
Then we have
$$
\eqalignno{
R(a_j,a_k) &= {q_j p_k - p_j q_k \over a_j-a_k}\, ,\ \ j\neq k,  &(1.23a) \cr
R(a_j,a_j) &=\pi^2 q_j^2 +p_j^2 \ + \sum_{k=1}^{2m} (-1)^k R(a_j,a_k)R(a_k,a_j)
(a_j-a_k)\, . &(1.23b)}
 $$

 Thus, the parameter dependence of $\det(I-\wh{K})$
is determined through the quantities $\{q_j, p_j\}_{j=1, \dots 2m}$.
Differentiating
equations (1.17a,b) with respect to the parameters $a_j$ and taking the
appropriate
limits, these may be shown to satisfy the following system of equations:
$$
\eqalign{ {\partial q_j \over \partial a_k} &= (-1)^k
R(a_j,a_k) q_k\, ,\ \ j\neq k \cr {\partial p_j \over \partial a_k} &= (-1)^k
R(a_j,a_k)
p_k\, ,\ \ j\neq k  \cr {\partial q_j\over \partial a_j} &= p_j - \sum_{k\neq
j}
(-1)^kR(a_j,a_k) q_k\cr {\partial p_j\over \partial a_j} &= -\pi^2 q_j -
\sum_{k\neq j}
(-1)^k R(a_j,a_k)p_k.} \eqno(1.24)
$$
This overdetermined system, which we refer to as the JMMS equations, is in fact
Frobenius integrable, and has a Hamiltonian structure which we summarize in the
next subsection.
\smallskip
\noindent {\it 1c. Hamiltonian Structure of the JMMS Equations}

For convenience, define a new normalization of the coordinates
$$
\eqalign{
q_{2j} &:=-{i\over 2}x_{2j}, \quad  q_{2j+1}:={1\over 2} x_{2j+1} \cr
p_{2j} &:=-iy_{2j}, \quad  p_{2j+1}:=y_{2j+1}} \eqno(1.25)
$$
and let
$$
G_j(x,y):={\pi^2\over 4} x_j^2+y_j^2-{1\over 4}\sum_{k=1\atop k\neq j}^{2m}
{(x_j y_k-x_k y_j)^2\over a_j-a_k}. \eqno(1.26)
$$
In terms of these quantities, we have
$$
\omega(a)=\sum_{j=1}^{2m} G_j(x,y)\, da_j.  \eqno(1.27)
$$
Defining canonical Poisson brackets:
$$
\left\lbrace x_j,x_k\right\rbrace = \left\lbrace y_j,y_k\right\rbrace=0, \quad
\left\lbrace x_j,y_k \right\rbrace = \delta_{jk}, \quad j, k =1, \dots, 2m,
\eqno(1.28)
$$
the system of equations (1.24) is seen to have the nonautonomous Hamiltonian
form
$$
 d_a x_j=\left\lbrace x_j,\omega(a)\right\rbrace \quad {\rm and} \quad
d_a y_j =\left\lbrace y_j,\omega(a)\right\rbrace.  \eqno(1.29)
$$
Moreover, the set of Hamiltonians $\{G_j\}_{j=1,\dots 2m}$ are easily verified
to be
generically functionally independent and in involution:
 $$
 \left\lbrace G_j,G_k\right\rbrace = 0 \quad{\rm for \ all}\quad
  j,k=1,\ldots,2m. \eqno(1.30)
$$
{}From this follows that the system (1.29), or equivalently (1.24), is
Frobenius
integrable, and that $\o$ is locally an exact differential:
$$
\o =d_a \,\log\,  \tau, \eqno(1.31)
$$
where the value of the tau-function $\tau$ along the integral curves coincides
with
$\log\left(\det(I-\wh{K})\right)$.

   The significance of this Hamiltonian structure in terms of spectral
invariants on
loop algebras will be explained in \S 2. First, another similar system
occurring in
the computation of level spacing distribution functions of scaled random
matrices at
the ``edge of the spectrum'' will be described. For the details
of what follows see {\bf [TW2]}.  (Here also only the case of
equal $\lambda_j$'s was considered, the generalization to the
case of unequal $\lambda_j$'s being straightforward.)
\smallskip
 \noindent {\it 1d. The Airy Kernel System and  Distributions at the Edge of
the
Spectrum: }

   Scaling  at
the
``edge of the spectrum'' {\bf [Mo]}, {\bf [Fo]},  {\bf [TW2]},
corresponds to choosing  $X_0\sim\pm\sqrt{2N}$
and gives
 rise to a different Fredholm
determinant, in which the sine kernel (1.8) is replaced by the {\it Airy
kernel}
$$\eqalign{
 K(x,y) :=& {A(x) A^\prime(y) - A^\prime(x) A(y) \over x-y}\cr
=&\lim_{N\rightarrow\infty} {1\over 2^{1/2} N^{1/6}}
K_N\left(\sqrt{2N}+{x\over 2^{1/2} N^{1/6}},\sqrt{2N}+{y\over 2^{1/2} N^{1/6}}
\right)\, ,\cr} \eqno(1.32)
$$
where now, $A(x)$ is an Airy function
$$
 A(x) = {\rm Ai}(x).   \eqno(1.33)
$$
The logarithmic differential of the Fredholm determinant (1.14) in this case is
also
given by formula (1.21) where, using the same notation as above, the resolvent
kernel
$R(x,y)$ is still of the form (1.19a,b), with the functions $Q(x,a), P(x,a)$
obtained by
replacing $A(x)$ in eq.~(1.17a,b), (1.18) by (1.33). Define $\{q_j, p_j\}_{j=1,
\dots
2m}$ as in (1.22a,b), and introduce two further quantities:
$$
\eqalignno{
u :=& \sum_{j=1}^m \l_j\int_{a_{2j-1}}^{a_{2j}} A(x) Q(x,a)\,  dx &(1.34a)\cr
v :=& \sum_{j=1}^m\l_j\int_{a_{2j-1}}^{a_{2j}} A(x) P(x,a)\,  dx.&(1.34b)}
$$
Then similarly to the previous case, we have,
$$
\eqalignno{
R(a_j,a_k) &= {q_j p_k - p_j q_k \over a_j-a_k}\, ,\ \ j\neq k   &(1.35a)\cr
R(a_j,a_j) &=\sum_{k\neq
j}(-1)^k\,  {(q_j p_k - p_j q_k)^2\over a_j - a_k}  + p_j^2 - a_j q_j^2 - 2 p_j
q_j u +
2 q_j^2 v,  &(1.35b)
}
$$
where all the limits are again taken within the intervals $J_j, J_k$.
Differentiating
with respect to the parameters $\{a_j\}_{j=1, \dots 2m}$ gives the system
$$
\eqalign{
{\partial q_j\over\partial a_k}&= (-1)^k\,  {q_j p_k - p_j q_k\over
a_j-a_k}\, q_k\,\ \ \ (j\neq k), \cr
{\partial p_j\over\partial a_k}&= (-1)^k\,  {q_j p_k - p_j q_k\over
a_j-a_k}\, p_k \ \ \  (j\neq k), \cr
{\partial q_j\over\partial a_j}&=
-\sum_{k\neq j} (-1)^k\,  {q_j p_k - p_j q_k \over a_j - a_k}\, q_k
+p_j - q_j u\, \cr
{\partial p_j\over\partial a_j}&=
-\sum_{k\neq j} (-1)^k\,  {q_j p_k - p_j q_k \over a_j - a_k}\, p_k
+a_j q_j + p_j u-2 q_j v   \cr
{\partial u\over\partial  a_j}&=(-1)^j q_j^2 \cr
{\partial v\over\partial  a_j}&= (-1)^j p_j q_j.} \eqno(1.36)
$$

   Again, introducing new coordinates $\{x_j, y_j\}_{j=0, \dots 2m}$ by:
$$
\eqalign{
q_{2j} &:=-{i\over 2}x_{2j}, \quad  q_{2j+1}:={1\over 2} x_{2j+1} \cr
p_{2j} &:=-iy_{2j}, \quad \ p_{2j+1}:=y_{2j+1}\cr
u & :=y_0, \qquad \qquad \ \ v :={1\over 2} x_0. } \eqno(1.37)
$$
and defining canonical Poisson brackets
$$
\left\lbrace x_j,x_k\right\rbrace = \left\lbrace y_j,y_k\right\rbrace=0, \quad
\left\lbrace x_j,y_k \right\rbrace = \delta_{jk}, \quad j, k =0, \dots, 2m,
\eqno(1.38)
$$
eqs.~(1.36) have the Hamiltonian form
$$
 d_a x_j=\left\lbrace x_j,\omega(a)\right\rbrace \quad {\rm and} \quad
d_a y_j =\left\lbrace y_j,\omega(a)\right\rbrace, \quad j =0 , \dots , 2m,
\eqno(1.39)
$$
where
$$
\omega(a) =  d_a \log\left(\det(I-\wh{K})\right) = \sum_{j=1}^{2m} G_j(x,y)\,
da_j  \eqno(1.40)
$$
and
$$
G_j(x,y) = y_j^2 - {1\over 4} a_j x_j^2 - x_j y_j y_0 +{1\over 4} x_j^2 x_0
-{1\over 4}\sum_{k=1\atop k\neq j}^{2m}
{\left(x_j y_k - x_k y_j\right)^2\over a_j - a_k}. \eqno(1.41)
$$
It is straightforward to verify that these $G_j$'s are also in involution,
implying
the Frobenius integrability of the system (1.40) and the existence of a
$\tau$-function as
in eq.~(1.31). However, in this case, the phase space is $2(m+1)$--dimensional,
so the
$G_j$'s alone do not form a complete set of commuting integrals. It is easily
verified,
though, that the following additional independent function
$$
G_0(x,y) := y_0^2 -x_0 -{1\over 4}\sum_{i=1}^{2m} x_i^2 \eqno(1.42)
$$
is also in involution with the $G_j$'s, thereby forming a complete set
$\{G_j\}_{j=0,
\dots 2m}$ of commuting integrals. Introducing an additional flow parameter
$a_0$
corresponding to the Hamiltonian $G_0$, and letting
$$
\tilde{\o}(\tilde{a}) := \o(a) + G_0da_0, \eqno(1.43)
$$
where
$$
\wt{a} := (a_0, \dots , a_{2m}), \eqno(1.44)
$$
we have the following extended system, which is also Frobenius integrable:
$$
 d_{\tilde a} x_j=\left\lbrace x_j,\tilde\omega(\tilde a)\right\rbrace \quad
{\rm and}
\quad d_{\tilde a} y_j =\left\lbrace y_j,\tilde\omega(\tilde a)\right\rbrace,
\quad j =0
, \dots , 2m.  \eqno(1.45)
$$
Moreover, since $G_0$ is not explicitly dependent on the parameters $a_j$ and
Poisson
commutes with all the $\{G_j\}_{j=1, \dots 2m}$, it is in fact a constant.
Since all
the quantities $u,v,q_j, p_j \ra 0$ as the $a_j \ra \infty$, this constant
value
is just $0$.
\bigbreak\bigbreak
\bigskip \noindent
{\bf 2. Classical $R$-Matrix  and Isomonodromy Formulation }
\medskip
 Hamiltonian systems of the type encountered above can be understood within a
general Lie
algebraic setting (cf. {\bf [AHP]},
{\bf [H]},  {\bf [RS]}, {\bf [S]}) which we now
summarize. Let
$\lg$ be the loop algebra of smooth (real or complex),
 $sl(r)$-valued functions $X(\lambda)$,
defined on a
circle $S^1$ centered at the origin in the complex $\l$-plane.
 Define the Ad-invariant scalar
product $<\ , \ >$ on $\lg$ by:
$$
 <X,\ Y> = {1\over 2 \pi i}\oint_{S^1}\rm
tr\left(X(\l)Y(\l)\right)\, d\l .  \eqno(2.1)
$$
Interpreting this as the evaluation of a linear form $X$ on elements $Y \in
\lg$
allows us to identify $\lg$ with an open, dense subspace of the dual space,
which will be denoted $\lg^*$. Under this identification, both the adjoint
and coadjoint actions of the corresponding loop group $\LG$ are given by
conjugation:
$$
\Ad_g:X\lmt gXg^{-1}, \qquad  \Ad^*_g:X\lmt gXg^{-1},
\qquad g\in \LG,\quad X \in\lg.      \eqno(2.2)
$$
As a linear space, $\lg$ may be decomposed into a direct sum
$$
\lg = \lgp + \lgm ,      \eqno(2.3)
$$
where $\lgp$, $\lgm$ are the subalgebras consisting of elements $X(\l)$ that
admit, respectively,  holomorphic extensions to the interior and exterior of
$S^1$, with the elements $X \in \lgm$ satisfying $X(\infty) =0$.
Under the identification $\lg \sim \lg^*$, we have
$$
\lgp^* \sim \lgm, \qquad \lgm^* \sim \lgp,       \eqno(2.4)
$$
where $\lg_{\pm}^* \ss \lg^*$ are identified as the annihilators of
$\lg_{\mp}$.

 Let
$$
\eqalign{
P_{\pm}:\lg &\lra \lg_{\pm} \cr
 P_{\pm}: X & \lmt X_{\pm}}      \eqno(2.5)
$$
be the projections to these two subspaces relative to the decomposition (2.3)
(determined, e.g., by splitting the Fourier series on $S^1$ into positive and
negative parts) and define the endomorphism $R:\lg \ra \lg$ as the
difference:
$$
R := P_+ - P_-.      \eqno(2.6)
$$
Then $R$ is a {\it classical R-matrix} {\bf[S]}, in the sense that the bracket
$[\ ,\ ]_R$ defined on $\lg$ by:
$$
[X,Y]_R:= {1\over2}[RX,Y] + {1\over2}[X,RY]       \eqno(2.6)
$$
is skew symmetric and satisfies the Jacobi identity, determining a
new Lie algebra structure on the same space as $\lg$, which we denote $\lg_R$.
The Lie Poisson bracket on $\lg_R^* \sim \lg^*$ associated to the Lie
bracket $[\ ,\ ]_R$ is:
$$
\{f, g\}\vert_X =<[df, dg]_R,X>        \eqno(2.7)
$$
for smooth functions $f,g$ on $\lg_R^*$ (with the usual identifications,
$df\vert_X,dg\vert_X \in \lg^* \sim \lg)$. The corresponding group
$\LG_R$ is identified with the direct product $\LG_+ \times \LG_-$, where
$\LG_{\pm}$ are the loop groups associated to $\lg_{\pm}$. The adjoint and
coadjoint actions are given by:
$$
\eqalign{
\Ad_R(g): (X_ + +X_-) &\lmt g_+X_+g_+^{-1} + g_-X_-g_-^{-1}  \cr
\Ad^*_R(g): (X_+ +X_-) &\lmt (g_-X_+g_-^{-1})_+ + (g_+X_-g_+^{-1})_-, \cr
X_{\pm} \in \lg_{\pm}, & \qquad   g_{\pm} \in \LG_{\pm}}       \eqno(2.8)
$$

Let $a:=\{a_i\}_{i=1}^n $ be a set of $n$ distinct real
(or complex) numbers
and $\k := \{k, k_i\}_{i=1}^n$ a set of $n+1$ nonnegative integers.
Define the subspace $\ga \ss \lg_R^*$ to consist of elements
$X=X_+ + X_-$ with $X_+\in \lgp$  a polynomial in $\l$ of degree $k$ and
$X_-\in \lgm$ a rational function of $\l$, with poles of degree $\{k_1,
\dots, k_n\}$ at the points $\{a_1, \dots, a_n\}$. Then $\ga \ss \lg_R^*$ is
$Ad_R^*$-invariant, and hence a Poisson subspace. The coefficient $Y$ of the
leading
term in the degree $k$ polynomial $X_+$ is $Ad_R^*$ invariant, and hence
constant under
all Hamiltonian flows. Let ${\cal I}_{a,\k} :=I(\lg^*)\vert_{\ga}$ be the ring
of
$Ad^*$-invariant polynomials on $\lg^* \sim\lg_R^*$,  restricted to $\ga$, and
define, for
$X \in \ga$, the $sl(r)$-valued polynomial function:
$$
L(\l) := a(\l) X(\l),        \eqno(2.9)
$$
where
$$
a(\l) := \prod_{i=1}^n (\l - a_i)^{k_i}.       \eqno(2.10)
$$
Then the ring $\Iak$ is generated by the coefficients of the
characteristic polynomial:
$$
{\cal P}(\l,z) :=\det(L(\l) -z I),       \eqno(2.11)
$$
and the characteristic equation
$$
 {\cal P}(\l,z) = 0       \eqno(2.11)
$$
defines, after suitable compactification (and possible desingularization), an
algebraic curve $\Gamma$, called the ``spectral curve'', which is generically
an
$r$-fold branched cover of ${\bf CP}^1$.
\bigbreak
It follows from the Adler-Kostant-Symes (AKS) theorem  that:
\item{1)} The ring $\Iak$ is Poisson commutative (with respect to
the $\{\ ,\ \}_R$ Poisson bracket).
\item{2)}  For a Hamiltonian $G\in {\cal I}_{a,\k}$, Hamilton's equations are
given by:
$$
{dX \over dt} = \{X,G\} = [dG_+,X] = -[dG_-, X]       \eqno(2.12)
$$
(The last equality following from the fact that $[dG,X]=0$, which is the
infinitesimal form of $\Ad^*$-invariance).
\smallskip
This means that the spectral curve $\Gamma$ is invariant under the flows
generated by the
elements of  $\Iak$, since any two coefficients of the spectral polynomial
${\cal P}(\l,z)$ Poisson commute.  It may also be shown {\bf[AHP]}, {\bf
[AHH1]} that,
for sufficiently generic initial conditions and coadjoint  orbits  ${\cal O}$
in
$\ga$, the elements of $\Iak$, restricted to ${\cal O} \ss \ga$,
define completely integrable Hamiltonian systems; that is, the number of
independent generators in $\Iak$ equals one half the dimension of the
orbit. Through a standard construction {\bf [AvM]}, {\bf [RS]}, {\bf [AHH1]},
{\bf [AHH2]}, the flows they generate are shown to linearize on the Jacobi
variety
associated to $\Gamma$.

We now specialize to the case $ \{k=1, k_i=1\}_{i=1,\dots n}$. Then
$X(\l)$ has the form $$
X(\l) = \l Y + N_0+ \sum_{i=1}^n {N_i \over \l - a_i}       \eqno(2.13)
$$
and the $\Ad_R^*$-action is:
$$
Ad_R^*(g_+,g_-): X(\l) \lmt \l Y + N_0 + [\gamma_0,Y] +
\sum_{i=1}^n{g_iN_ig_i^{-1}\over \l -a_i},        \eqno(2.14)
$$
where
$$
\gamma_0 :={1\over 2\pi i}\oint_{S^1}g_-(\l)\, d\l,  \qquad g_i := g_+(a_i),
\qquad i=1,\dots n.        \eqno(2.15)
$$
Hence, each coadjoint orbit in $\ga$ consists of elements of the form:
$$
X(\l) = \l Y + C + [\gamma_0,Y] + \sum_{i=1}^n {N_i \over \l - a_i},
 \eqno(2.16)
$$
where
$$
N_0 := C + [\gamma_0,Y], \eqno(2.17)
$$
with $Y$ and $C$  constant matrices, and the $N_i$'s have fixed Jordan
normal form.

Let
$$
{1\over 2} \tr X^2 = y\l^2 + c \l + G_0 + \sum_{i=1}^n {G_i \over \l -a_i}
 + \sum_{i=1}^n {K_i \over (\l -a_i)^2},
\eqno(2.18)
$$
where
$$
y := {1\over 2}\tr Y^2, \qquad c:= \tr (YN_0) =\tr(Y C)    \eqno(2.19)
$$
and

$$
\eqalign{
G_0 &:= {1\over 4\pi i} \oint_{S^1} \tr X^2(\l) {d\l \over \l} =
\tr (Y\sum_{i=1}^n N_i)  + {1\over 2}\tr N_0^2 \cr
G_i &:= {1\over 4\pi i} \oint_{\l=a_i} \tr X^2(\l)\,  d\l = a_i \tr (Y N_i) +
\tr (N_0 N_i) + \sum_{j\neq i}^n{\tr (N_i N_j) \over a_i -a_j},
\cr
K_i &:= {1\over 4\pi i} \oint_{\l=a_i}(\l - a_i) \tr X^2(\l) \,  d\l
={1\over 2} \tr (N_i^2),   \cr
 i&=1, \dots  n}
\eqno(2.20)
$$
with the integrals $\oint_{\l=a_i}$ taken around a small circle enclosing only
this pole.
Then $y$, $c$ and $\{K_i\}_{i=1,\dots n}$ are all Casimir invariants
and hence constant on each $\Ad_R^*$ orbit, and $\{G_i\}_{i=0,\dots n}$ are
independent
elements of the Poisson commuting ring $\Iak$.  Denoting the time parameters
for the Hamiltonian flows generated by $\{G_i\}_{i=0,\dots n}$ as
$\{t_i\}_{i=0,\dots n}$,
the AKS theorem implies that Hamilton's equations are of the form:
$$
\eqalign{
{\di X\over \di t_0} &= \{X, G_0\} = [(dG_0)_{+}, X]  \cr
{\di X\over \di t_i} &= \{X, G_i\} = -[(dG_i)_{-}, X], \qquad i=1, \dots n}
\eqno(2.21)
$$
where
$$
(dG_0)_{+}= Y, \qquad  (dG_i)_{-}= -{N_i \over \l -a_i}, \qquad i=1, \dots n .
\eqno(2.22)
$$

Evaluating residues at $\{\l =a_i\}_{i=1, \dots n}$ and $\l = \infty$, we see
that this is equivalent to:
$$
\eqalign{
{\di N_i\over \di t_0}&=[Y, N_i], \qquad
{\di N_0\over \di t_0} =[Y, N_0], \qquad {\di N_0\over \di t_i} =[Y,N_i],  \cr
{\di N_i\over \di t_j}&= {[N_i,N_j]\over a_i - a_j}, \quad i \neq j,
 \qquad
{\di N_i\over \di t_i} = [a_iY + N_0 +\sum_{j\neq i}{N_j \over a_i - a_j},
\ N_i] \cr
i,j& = 1, \dots n .}   \eqno(2.23)
$$

If we now reinterpret these as nonautonomous Hamiltonian systems, by
identifying the constants $\{a_i\}$ with the flow parameters
$\{t_i =a_i\}_{i=1,\dots n}$, system (2.22) becomes (cf. {\bf [JMMS]}, {\bf
[JMU]}):
$$
\eqalign{
{\di N_i\over \di a_0}&=[Y, N_i], \qquad
{\di N_0\over \di a_0}=[Y, N_0], \qquad {\di N_0\over \di a_i} =[Y,N_i],  \cr
{\di N_i\over \di a_j}&= {[N_i,N_j]\over a_i - a_j}, \quad i \neq j,
 \qquad
{\di N_i\over \di a_i} = [a_iY + N_0 +\sum_{j\neq i}{N_j \over a_i - a_j},
\ N_i] \cr
i,j& = 1, \dots n .}   \eqno(2.24)
$$
(Here $t_0$ has similarly been renamed $a_0$, but it does not appear as a
parameter in the system, and hence represents autonomous flow.)
This system is equivalent to the commutativity of the following system of
differential operators:
$$
\eqalign{
{\cal D}_{\lambda} &:= {\di \over \di \lambda} -\lambda Y -N_0 -
 \sum_{i=1}^n {N_i \over \lambda - a_i}, \cr
 {\cal D}_{i}&:= {\di \over \di a_i} + {N_i \over \lambda - a_i}, \cr
{\cal D}_{0}&:= {\di \over \di a_0}  -  Y.}  \eqno(2.25)
$$
It follows that the monodromy of the operator ${\cal D}_{\lambda}$ is
preserved under the deformations parametrized by $\lbrace a_0,a_1,\ldots,
a_n\rbrace$.
\par
To obtain the system (1.26)--(1.29), we take
the $\lbrace a_j \rbrace_{j=1}^n$ as real,
 $r=2$, $n=2m$,  $Y=0$ (so really,
we are
in the Poisson subspace with $k=0$), and choose
$$
C = \pmatrix {0 & {\pi^2\over 2}\cr -2 & 0}     \eqno(2.26)
$$
and the  $N_i$'s with Jordan normal form $\pmatrix {0 & 0\cr 1 & 0}$.
The $\Ad_R^*$ orbit ${{\cal O}_0} \ss \ga$ then consists of elements of
the form:
$$
X(\l) = \pmatrix {0 & {\pi^2\over 2}\cr -2 & 0} +
{1\over 2} \sum_{i=1}^{2m}{\pmatrix {-x_iy_i & -y_i^2 \cr x_i^2 & x_iy_i}
 \over \lambda - a_i} .  \eqno(2.27)
$$
Here, we have parametrized the orbit by elements of
 $M_0:= {\bf R}^{4m}/({\bf Z}_2)^{4m}$, where
\vskip -7pt \noindent $((x_1, \dots, x_{2m}),(y_1,
\dots, y_{2m})) \in {\bf R}^{4m}$, and the  $({\bf Z}_2)^{4m}$ action is
generated by
reflections in the coordinate hyperplanes:
$$
\eqalign {
(\eps_1, \dots, \eps_{2m}): ((x_1, \dots, x_{2m}),(y_1, \dots, y_{2m}))
&\lmt  ((\eps_1x_1, \dots, \eps_{2m}x_{2m}),(\eps_1y_1, \dots,
\eps_{2m}y_{2m}))
\cr \{\eps_i &= \pm 1\}_{i=1, \dots 2m}.}
 \eqno(2.28)
$$
The map $ J_0: M_0 \lra \lg_R^*$ defined by
$$
J_0:(( x_1, \dots, x_{2m}), ( y_1, \dots, y_{2m})
\lmt \pmatrix {0 & {\pi^2\over 2}\cr -2 & 0} +
{1\over 2} \sum_{i=1}^{2m}{\pmatrix {-x_iy_i & -y_i^2 \cr x_i^2 & x_iy_i}
 \over \lambda - a_i}     \eqno(2.29)
$$
is a symplectic embedding with respect to the standard symplectic
structure:
$$
\omega = \sum_{i=1}^{2m} dx_i \wedge dy_i
\eqno(2.30)
$$
on $M_0$ and the orbital (Lie-Kostant-Kirillov) symplectic structure on ${\cal
O}_0$.
(This is a special case of the moment map embeddings into rational Poisson
subspaces of the dual of loop algebras developed in {\bf [AHP]}, {\bf [AHH3]}.)
Evaluating the Hamiltonians $\{G_i\}_{i=1, \dots 2m}$ defined in (2.20) for
this case
gives those of eqs.~(1.26) and the  $\{a_j\}$, $j\ne 0$ components of
the equations of motion (2.24) are equivalent to (1.29), up to quotienting by
the
$({\bf Z}_2)^{4m}$ action (2.28).

  To obtain the system (1.40)--(1.45), we again take $r=2$, $n=2m$,  but choose
$$
 Y=\pmatrix {0 &  -{1\over 2} \cr 0 & 0}, \quad
C = \pmatrix {0 & 0\cr -2 & 0},\quad
[\gamma_0,Y]= \pmatrix {y_0 & {x_0 \over 2} \cr 0 & -y_0},     \eqno(2.31)
$$
with the  $N_i$'s again of Jordan normal form $\pmatrix {0 & 0\cr 1 & 0}$.
In this case, the $\Ad_R^*$ orbit ${\cal O}_1 \ss \ga$ consists of elements of
the form:
$$
X(\l) = \pmatrix {y_0 & -{\lambda \over 2} + {x_0 \over 2}\cr -2 & -y_0}+
{1\over 2} \sum_{i=1}^{2m}{\pmatrix {-x_iy_i & -y_i^2 \cr x_i^2 & x_iy_i}
 \over \lambda - a_i} .  \eqno(2.32)
$$
Here, we have parametrized the orbit by elements of
 $M_1:= {\bf R}^2 \times {\bf R}^{4m}/({\bf Z}_2)^{4m}$, where
$(x_0,y_0) \in {\bf R}^2$, $((x_1, \dots, x_{2m}),(y_1, \dots, y_{2m})) \in
{\bf
R}^{4m}$, and the  $({\bf Z}_2)^{4m}$ action is again given by (2.28).
The map $ J_1: M_1 \lra \lg_R^*$ defined by
$$
J_1:((x_0, x_1, \dots, x_{2m}), (y_0, y_1, \dots, y_{2m})
\lmt \pmatrix {y_0 & -{\lambda\over 2} + {x_0 \over 2}\cr -2 & -y_0}
+
{1\over 2} \sum_{i=1}^{2m}{\pmatrix {-x_iy_i & -y_i^2 \cr x_i^2 & x_iy_i}
 \over \lambda - a_i}     \eqno(2.33)
$$
is  again a symplectic embedding, now with respect to the symplectic
structure:
$$
\omega = dx_0\wedge dy_0 + \sum_{i=1}^{2m} dx_i \wedge dy_i
\eqno(2.34)
$$
on $M_1$ and the orbital (Lie-Kostant-Kirillov) symplectic structure on ${\cal
O}_1$.
Evaluating the Hamiltonians $\{G_i\}_{i=0, \dots 2m}$ defined in (2.20) for
this case
gives those of eqs.~(1.41), (1.42) and the equations of motion (2.24) are
equivalent to
(1.45) -- again, up to quotienting by the $({\bf Z}_2)^{4m}$ action (2.28).

For this case, eqs.~(2.19), (2.31), (2.33) give $y=0$, $c=1$ and $\tr
(N_i^2)=0$, so
$$
-\det(X)= {1\over 2}\tr(X^2)= \l + G_0 +
\sum_{i=1}^{2m}{G_i \over \l -a_i} := {P(\l)\over a(\l)}, \eqno(2.35)
$$
where  $P(\l)$ is a monic polynomial of degree $2m+1$
$$
P(\l) = \l^{2m+1} + P_{2m}\l^{2m} + \dots + P_0     \eqno(2.36)
$$
with
$$
P_{2m} = G_0 -\sum_{i=1}^{2m}a_i, \qquad P_{2m -1} = \sum_{i=1}^{2m} G_i
-G_0 \sum_{i=1}^{2m} a_i,  \quad {\rm etc.}    \eqno(2.37)
$$
{}From eqs.~(2.9)--(2.11), we have the following equation for the spectral
curve  $\Gamma$:
$$
{\cal P}(\l, z) = z^2 - a(\l)P(\l) = 0,   \eqno(2.38)
$$
which shows that it is hyperelliptic, of genus $g=2m$, with $2m+1$ of its
branch points located at $\{a_1, \dots a_{2m}, \infty\}$ and the
others determined by the values of the roots of $P(\l)$.
For the system
(1.26)--(1.29) associated to the sine kernel, the leading term in (2.18)
vanishes, so the polynomial $P(\lambda)$ is of degree $2m$ and the curve has
genus $g=2m-1$, with no branch points at $\infty$.
For the autonomous
system (2.23), these curves are invariant under the flows.
 For the nonautonomous
system (2.24), which in these  cases  reduces to (1.29) or (1.45),
 it would be of interest to
determine the dependence of the spectral curves  on the deformation
parameters $\{a_i\}_{i=0,\dots 2m}$ through a system of PDE's involving the
spectral
invariants $\{G_i\}_{i=0, \dots 2m}$ alone. For the single finite interval
case, this is given in {\bf [JMMS]}, {\bf [TW1]} by the $\tau$-function form
of the
Painlev\'e equation ${P_V}$, while for the single semi-infinite
interval  case,
this is
given in {\bf [TW2]} by the corresponding $\tau$--function form of $P_{II}$.
\bigskip
\noindent{\bf Acknowledgments}
\par
It is a pleasure to acknowledge E.~L.~Basor,
F.~J.~Dyson, P.~J.~Forrester, A.~Its, and
M.~L.~Mehta for their helpful comments and encouragement and for sending us
their preprints prior to publication.
The second author wishes to thank the organizers of the program
 ``Low Dimensional
Topology and Quantum Field Theory'' for their kind hospitality at the
Isaac Newton Institute for Mathematical Sciences.  Particular thanks to
R.~J.\  Baxter and H.~Osborn for making the stay at Cambridge most pleasant.
\vfill\eject
 \centerline{\bf References}
\bigskip {\smalltype
\item{\bf [AHH1]} Adams, M.~R., Harnad, J.\  and Hurtubise, J., ``Isospectral
Hamiltonian Flows in Finite and Infinite Dimensions II.  Integration of
Flows,''
 {\it Commun.\  Math.\  Phys.\/} {\bf 134}, 555--585 (1990).
\item{\bf [AHH2]}  Adams, M.~R., Harnad,~J. and  Hurtubise,~J., ``Darboux
Coordinates and Liouville-Arnold Integration  in Loop Algebras,''   preprint
CRM  (1992) (to appear in {\it Commun.\  Math.\  Phys.\/} (1993))
.%
\item{\bf [AHH3]} Adams, M.~R., Harnad, J.\  and Hurtubise, J.,
``Dual Moment Maps to Loop Algebras,''  {\it Lett.\  Math.\  Phys.\/} {\bf 20},
 294--308 (1990).
\item{\bf [AHP]} Adams, M.~R., Harnad, J.\  and Previato, E., ``Isospectral
Hamiltonian Flows in Finite and Infinite Dimensions I. Generalised Moser
Systems
and Moment Maps into Loop Algebras,''   {\it Commun.\ Math.\  Phys.\/} {\bf
117},
451--500 (1988).
 \item{\bf[AvM]} Adler, M.\ and van Moerbeke, P., ``Completely Integrable
Systems, Euclidean Lie Algebras, and Curves,''  {\it Adv.\  Math.\/}
{\bf 38}, 267--317 (1980); ``Linearization of Hamiltonian Systems, Jacobi
Varieties and Representation Theory,''  {\it ibid.} {\bf 38}, 318--379 (1980).
\item{\bf[BTW]} Basor, E.~L., Tracy, C.~A., and Widom, H.,
``Asymptotics of Level-Spacing Distributions for Random Matrices,''
{\it Phys.\ Rev.\ Letts.\/} {\bf 69}, 5--8 (1992).

\item{\bf [Dy]} Dyson, F.~J., ``The Coulomb Fluid and the Fifth Painlev{\'e}
Transcendent,'' IASSNSS-HEP-92/43 preprint, to appear in the proceedings
of a conference in honor of C.~N.~Yang, ed.\ S.-T.~Yau.
\item{\bf [Fo]} Forrester, P.~J., ``The Spectrum Edge of Random Matrix
Ensembles,'' preprint.
\item{\bf [H]} Harnad, J., ``Dual Isomonodromy Deformations and Moment Maps to
Loop
Algebras,''  preprint CRM-1844 (1993).
\item{\bf [IIKS]} Its, A.~R., Izergin, A.~G., Korepin, V.~E.\  and Slavnov,
N.~A.,
``Differential Equations for Quantum Correlation Functions,''
{\it Int.~J.~Mod.~Phys.\/}  {\bf
B4}, 1003--1037 (1990).
\item{\bf [IIKV]} Its, A.~R., Izergin, A.~G., Korepin, V.~E.\  and Varzugin,
G.~G.,
``Large Time and Distance Asymptotics of Field Correlation Function of
Impenetrable
Bosons at Finite Temperature,''  {\it Physica\/}  {\bf 54D}, 351--395 (1992).
 \item{\bf[JMMS]} Jimbo, M., Miwa, T., M{\^ o}ri, Y. and Sato, M.,, ``Density
Matrix of an Impenetrable Bose Gas and the Fifth Painlev\'e Transcendent'',
{\it Physica\/} {\bf 1D}, 80--158 (1980).
 \item{\bf[JMU]} Jimbo, M., Miwa, T.,and  Ueno, K., ``Monodromy Preserving
Deformation of Linear Ordinary Differential Equations with Rational
Coeefficients I.,''  {\it Physica\/} {\bf 2D}, 306--352 (1981).
 \item{\bf[JM]} Jimbo, M., and Miwa, T., ``Monodromy Preserving
Deformation of Linear Ordinary Differential Equations with Rational
Coeefficients II, III.,''  {\it Physica\/} {\bf 2D}, 407--448 (1981); {\it
ibid.},
{\bf 4D}, 26--46 (1981).
\item{\bf[Mc]} McCoy, B.~M., ``Spin Systems, Statistical Mechanics and
Painlev{\'e}
Functions,'' in {\it Painlev{\'e} Transcendents: Their Asymptotics and Physical
Applications\/}, eds.\ D.~Levi and P.~Winternitz, Plenum Press, New York
(1992),
pgs.\ 377--391.
\item{\bf[Me1]} Mehta, M.~L., {\it Random Matrices\/}, 2nd ed.,  Academic
Press,
San Diego,
(1991).
\item{\bf[Me2]} Mehta, M.~L.,  ``A Non-linear Differential
Equation and a Fredholm Determinant,''  {\it J.\ de Phys.\  I\/}{\bf
2},1721--1729
(1992).
\item{\bf [MM]} Mehta, M.~L.\ and Mahoux, G., ``Level Spacing Functions and
Non-linear Differential Equations,'' SPh-T/92--107 preprint.
\item{\bf [Mo]} Moore, G., ``Matrix Models of 2D Gravity and Isomonodromic
Deformation,''  {\it Prog.\ Theor.\ Phys.\ Suppl.\/}  No.\  {\bf 102}, 255--285
(1990).
\item{\bf [MTW]} McCoy, B.~M., Tracy, C.~A., and  Wu, T.~T., ``Painlev{\'e}
Functions
of the Third Kind,'' {\it J.\ Math.\ Phys.\/} {\bf 18}, 1058--1092 (1977).
\item{\bf [Po]} Porter, C.~E., {\it Statistical Theory of Spectra:
Fluctuations\/},
Academic Press, New York (1965).
 \item{\bf [RS]} Reiman, A.~G., and Semenov-Tian-Shansky,
M.A., ``Reduction of Hamiltonian systems, Affine Lie algebras and Lax Equations
I,
II,''  {\it Invent.\  Math.\/} {\bf 54}, 81--100 (1979);  {\it ibid.}  {\bf
63},
423--432 (1981).
\item{\bf [S]} Semenov-Tian-Shansky, M.~A.,
``What is a Classical R-Matrix,''
{\it Funct.\  Anal.\  Appl.\/} {\bf 17} 259--272 (1983);
``Dressing Transformations and Poisson Group Actions,''
{\it Publ.\  RIMS Kyoto Univ.\/} {\bf 21} 1237--1260 (1985).
\item{\bf [SMJ]} Sato, M., Miwa, T., and Jimbo, M., ``Holonomic Quantum Fields
I--V,''{\it Publ.\  RIMS Kyoto Univ.\/}
{\bf 14}, 223--267 (1978); {\it ibid.} {\bf 15},
 201--278 (1979); {\it ibid.} {\bf 15}, 577--629
(1979);
{\it ibid.}{\bf 15}, 871--972 (1979); {\it ibid.} {\bf 16}, 531--584 (1980).
\item{\bf [TW1]} Tracy, C.~A., and Widom, H., ``Introduction to Random
Matrices,''   UCD preprint ITD 92/93--10 (1992), to appear in {\it VIIIth
Scheveningen
Conf.\  Proc.\/}, Springer Lecture Notes in Physics.
\item{\bf [TW2]} Tracy, C.~A., and Widom, H., ``Level Spacing Distributions
and the Airy Kernel,''   UCD preprint ITD 92/93--9 (1992).
\item{\bf [W]} Widom, H., ``The Asymptotics of a Continuous Analogue
of Orthogonal Polynomials,'' to appear in J.\ Approx.\ Th.
\item{\bf[WMTB]} Wu, T.~T., McCoy, B.~M., Tracy, C.~A.,
and  Barouch, E., ``Spin-Spin
Correlation Functions for the Two-Dimensional Ising Model: Exact Theory
in the Scaling Region,'' {\it Phys.\ Rev.\/}  {\bf B13}, 316--374 (1976).
\item {}}
\vfill \eject

\end